\documentclass[11pt]{article}
\usepackage[latin9]{inputenc}
\usepackage{amsmath}
\usepackage{esint}

\makeatletter
\usepackage{jheppub}

\pdfoutput=1

\usepackage{mathrsfs}\usepackage{epstopdf}\usepackage{slashed}\usepackage{color}\usepackage{tikz}
\usetikzlibrary{matrix}
\usetikzlibrary{positioning}

\title{No-Go Theorems on Localization of Gravity around Higher Codimensional Branes in Noncompact Extra Dimensions}
\author{Shing Yan Li}

\affiliation{Center for Theoretical Physics, Department of Physics, Massachusetts Institute of Technology\\
77 Massachusetts Ave, Cambridge, MA 02139, USA}
\affiliation{Perimeter Institute for Theoretical Physics\\31 Caroline St.
N, Waterloo ON, Canada, N2L 2Y5}
\affiliation{Department of Physics, Hong Kong University of Science and Technology\\
Clear Water Bay, Kowloon, Hong Kong}

\emailAdd{sykobeli@mit.edu}

\abstract{We study the brane world scenario of a single brane (or a single stack of branes) with codimension higher
than one. When the extra dimensions are not small, localization of
gravity around the brane is needed in order to reproduce the observable
four-dimensional gravity. We focus on the case of noncompact extra
dimensions, where the possibility of localized gravity becomes non-trivial.
We show that in large class of gravity models, localization of massless gravity is not
possible for codimension-2 branes with at least one noncompact
extra dimension. With additional mild assumptions on field backgrounds, 
we also show that it is not possible for higher codimensional 
branes with two or more noncompact extra dimensions.}

\preprint{MIT-CTP/5262}

\makeatother

\begin{document}
\maketitle \flushbottom

\section{Introduction\label{sec:Introduction}}

Extra dimensions are well-motivated by, for example, string theory
and have rich and novel impacts on phenomenology, such as providing
potential solutions to hierarchy problems in particle physics and
cosmology. Some well-known examples are the cosmological constant
problem \cite{Burgess:1999qw,ArkaniHamed:2000eg,Chen:2000at,Kachru:2000hf,Dimopoulos:2001ui,Aghababaie:2003wz,Burgess:2003bia,Carroll:2003db,Navarro:2003vw,Burgess:2004ib,Burgess:2004kd,Burgess:2011va,Charmousis:2017rof}
and the Higgs hierarchy problem \cite{Randall:1999ee,Gherghetta:2000qt,Scrucca:2003ut,Dudas:2005gi,Sakamoto:2006wf,Burgess:2008ka}.
On the other hand, mechanisms are needed to implement a noncompact
four-dimensional spacetime into the extra dimensions, in order to
be consistent with our observations. So far there are two well-known
ways, namely compactification and the brane world scenario.

In the former case, the extra dimensions are ``packed'' into tiny
compact space, such that they are too small to be seen within the
energy scale we can achieve. In the latter case, a space-filling 3-brane
sits in the extra dimensions called bulk, and our universe is trapped
on the brane. While the extra dimensions must be small in compactification,
those in the brane world scenario can be large or even infinitely
large i.e. noncompact. In this paper, we investigate how noncompact
extra dimensions can affect the low-energy physics, in order to judge
the validity of these theories.

An essential feature of such theories is localization of gravity,
which can be explained as follows. The matter (Standard Model) particles
can be naturally trapped onto the brane, such as in the context of
D-branes. Therefore these particles can ``see'' fewer dimensions.
However, gravity is dynamics of spacetime, and can freely propagate
through the whole spacetime. Therefore to recover the observed Newton's
law, which implies four-dimensional gravity, mechanisms of localization
of gravity is needed to build the brane world scenario. For simplicity,
we always focus on localized massless gravity only, although massive
gravity is not completely ruled out by phenomenology. As we will see,
it is always possible in large but compact extra dimensions, while
it is hard in noncompact extra dimensions due to non-trivial boundary
conditions. Some aspects of localized gravity with higher codimensions
are already studied in literature such as in \cite{Csaki:2000fc,Oda:2001ss,Olasagasti:2003fw,Bachas:2011xa},
but most of them focused on compact extra dimensions only, and did
not have a concrete conclusion on the case of noncompact extra dimensions.

For codimension-1 brane, localization of gravity is easily achieved
in the Randall-Sundrum model \cite{Randall:1999vf}, where the four-dimensional
massless graviton is localized by a warp factor induced by fine-tuning
between positive brane tension and negative bulk cosmological constant.
It is later generalized to the codimension-1 Karch-Randall model \cite{Karch:2000ct}.
It is then natural to ask whether it is also possible with higher
codimensions. It is non-trivial because they have distinct dynamics.
Codimension-1 sources usually cause a jump of fields at brane position,
determined by the Israel junction conditions \cite{Israel:1966rt}.
However, higher codimensional sources generically cause divergences
at brane position. Such behavior is similar to the Coulomb potential
in dimension higher than one, which divergences are caused at charge
positions. This is why the dynamics of codimension-1 brane are much
more well studied in literature. Techniques to regularize these divergences
for codimension-2 branes are developed in \cite{Burgess:2007vi,Burgess:2008yx,Bayntun:2009im},
and it is straightforward to generalize them to even higher codimensions.

In this paper, we show that generically there is no localized gravity
around codimension-2 branes with at least one noncompact extra dimension,
or higher codimensional branes with at least two noncompact extra
dimensions. We consider a general theory with Einstein gravity, form
field backgrounds and dilaton, which includes large class of supergravity
models. We prove that in generic background, the warp factor far from
the brane cannot vanish asymptotically, thus cannot localize or normalize
the massless graviton mode function. We also derive the brane-bulk
boundary conditions which limit the form of the mode function, implying
that localized gravity can only be achieved by warp factors. Hence
localization of gravity is not possible. To achieve so, we use techniques
from analysis to study the general features of the field equations
without solving them. As a corollary, we also show that when a codimension-1
localized gravity model actually has one compactified hidden dimension,
within our model the brane must also wrap that dimension. Note that
we only consider a single brane (or a single stack of branes). This
simplifies the calculation much, and excludes more complicated setups
of higher codimensions such as brane intersections.

This paper is organized as follows. In Section \ref{sec:Codimension-2-Branes},
we use codimension-2 branes as example to study the dynamics among
branes, noncompact extra dimensions and massless graviton modes. We
review a general setup of such models, which can be easily generalized
into higher codimensions. We then examine the boundary conditions
to determine the consistent way to localize gravity, which is by warp
factor. After that, we prove the no-go theorem for codimension-2 branes.
We prove it separately for the cases of one and two noncompact extra
dimensions. This includes the case where a Randall-Sundrum-like solution
with a codimension-2 brane seems to trivially appear. We show that
such solution does not exist.

In Section \ref{sec:Branes-with-Higher}, we study localized gravity
around higher codimensional branes in general. To build motivation
to study such no-go theorems from some realistic models, we first
use a (infinitely long) Klebanov-Strassler throat \cite{Klebanov:2000hb,Klebanov:2000nc},
which is a well-known string theory solution, as an example. We then
generalize the setup of the system and prove the no-go theorem for
higher codimensions. To achieve so, additional assumptions on the
background must be made. Namely, the background is non-oscillating,
and under presence of form fields with nonvanishing couplings or brane
curvature.

In Section \ref{sec:Conclusion}, we conclude and add some remarks
to our results.

\section{Codimension-2 Branes\label{sec:Codimension-2-Branes}}

In this section, we use codimension-2 branes as example to demonstrate
general features of the dynamics of brane-bulk system and gravitons.
From these we prove the no-go theorems for codimension-2 branes.

\subsection{General Setup\label{subsec:General-Setup}}

\subsubsection*{Brane-Bulk System}

We first establish the brane-bulk system, which is most reviewed in
\cite{Bayntun:2009im}. For the bulk, we consider a simple model of
scalar-Einstein-Maxwell system with single fields. Similar but more
complicated arguments should hold for the cases of multiple fields,
so the model below is general enough to describe the bosonic parts
of large class of supergravity models. To implement the 4D universe,
there is a maximally symmetric 3-brane sitting in the bulk. The action
(in Einstein frame) is given by\footnote{Our metric is mostly plus, with Weinberg's curvature conventions \cite{Weinberg:1972kfs},
which differ from those of MTW \cite{Misner:1974qy} only by an overall
sign in the definition of the Riemann tensor.}

\begin{align}
S & =S_{bulk}+S_{brane}\,,\\
S_{bulk} & =-\int d^{4}x\,d^{2}y\,\sqrt{-g}\left(\frac{1}{2\kappa^{2}}R+\frac{1}{2}\partial^{M}\phi\partial_{M}\phi+\frac{1}{4}c\left(\phi\right)F_{MN}F^{MN}+V\left(\phi\right)\right)+S_{GH}\,,\\
S_{brane} & =-\int_{y_{b}}d^{4}x\,\sqrt{-\bar{g}}\,T_{b}\,,
\end{align}

\noindent where $g$ is the 6D background metric, $\bar{g}$ is the
unwarped 4D background metric, $R$ is the 6D Ricci scalar, $\phi$
is a scalar field, $F=dA$ is the Maxwell field strength, $c>0$ is
the field-dependent inversed coupling of the Maxwell field, $V$ is
the bulk potential including a bulk cosmological constant and the
scalar field potential, $T_{b}$ is the warped brane potential and
$y_{b}$ is the position of the brane. $S_{GH}$ is the Gibbons-Hawking
action at the brane-bulk boundary, added to restore the usual Einstein
equations. We first consider two noncompact extra dimensions, since
we will generalize such setup to higher codimensions. For a single
brane, the bulk geometry is rotationally invariant. We also include
warped geometries. Therefore the metric $g$ can be written as

\begin{equation}
ds^{2}=g_{MN}dx^{M}dx^{N}=e^{2A\left(r\right)}\bar{g}_{\mu\nu}dx^{\mu}dx^{\nu}+dr^{2}+e^{2B\left(r\right)}d\theta^{2}\,,
\end{equation}

\noindent where $A$ and $B$ are the warp factors which only depend
on $r$ and $\bar{g}$ is the maximally symmetric metric. In terms
of static coordinates,

\begin{equation}
\bar{g}_{\mu\nu}dx^{\mu}dx^{\nu}=-\left(1-k\rho^{2}\right)dt^{2}+\frac{d\rho^{2}}{1-k\rho^{2}}+\rho^{2}\left(d\varphi_{1}^{2}+\sin^{2}\varphi_{1}d\varphi_{2}^{2}\right)\,.
\end{equation}

\noindent The parameter $k=\left\{ -1,0,1\right\} $ corresponds to
unit-radius AdS, flat and dS spacetime respectively. To obtain two
noncompact extra dimensions, $r$ should go from zero to infinity,
and the warp factor $B$ should satisfy $e^{B}\rightarrow+\infty$
as $r\rightarrow\infty$. It is then natural to assume that $B$ is
increasing at large $r$. $B$ should also satisfy $\underset{r\rightarrow0}{\lim}\,e^{B}=0$,
in order to put the codimension-2 brane at $r=0$. From now on we
use capital English letter to denote all coordinates, and Greek letters
to denote 4D coordinates. Now we can choose a gauge such that the
only non-vanishing components of $F$ are

\begin{equation}
F_{r\theta}=-F_{\theta r}=A'_{\theta}\left(r\right)\,.
\end{equation}

\noindent By symmetry $A_{\theta}$ and $\phi$ also depend on $r$
only. Also from symmetry $T_{b}$ should be function of $\phi$ and
$A_{\theta}$ only. Therefore we perform a derivative expansion to
get

\begin{equation}
T_{b}=\tau_{b}\left(\phi\right)-\frac{1}{2}\Phi_{b}\left(\phi\right)\epsilon^{mn}F_{mn}+...\,,
\end{equation}

\noindent where $\tau_{b}$ is the warped brane tension and $\Phi_{b}$
is the warped localized magnetic flux on the brane \cite{Burgess:2011mt}.
The labels $m,n$ are for $r,\theta$ and $\epsilon_{mn}$ is the
corresponding Levi-Civita tensor.

Now for $r>0$, the bulk field equations are

\begin{equation}
\phi''+\left(B'+4A'\right)\phi'=\frac{\partial V}{\partial\phi}+\frac{1}{4}\frac{\partial c}{\partial\phi}F_{MN}F^{MN}\,,\label{scalarFieldEqn}
\end{equation}

\begin{equation}
R_{MN}-\frac{1}{2}Rg_{MN}=-c\kappa^{2}\left(F_{MP}\mathit{F_{N}}^{P}-\frac{1}{4}g_{MN}F_{PQ}F^{PQ}\right)-\kappa^{2}\left(\partial_{M}\phi\partial_{N}\phi-g_{MN}\left(\frac{1}{2}\partial^{P}\phi\partial_{P}\phi+V\right)\right)\,,
\end{equation}

\begin{equation}
\nabla_{M}\left(cF^{MN}\right)=0\,.
\end{equation}

\noindent Using the above ansatz, we get

\begin{equation}
\left(ce^{-B+4A}A'_{\theta}\right)'=0\Rightarrow A'_{\theta}=\frac{F_{0}}{c}e^{B-4A}\,,
\end{equation}

\noindent where $F_{0}$ is an integration constant and ``prime''
is partial derivative with respect to $r$. We further denote $f=\frac{\kappa^{2}F_{0}^{2}}{2c}$.
It is then straightforward to compute the components of the Einstein
equation:

\begin{equation}
6A'^{2}+3A'B'+B'^{2}+3A''+B''+fe^{-8A}-3ke^{-2A}+\kappa^{2}\left(\frac{1}{2}\phi'^{2}+V\right)=0\,,\quad\left(\mu\nu\right)\label{einstein4d}
\end{equation}

\begin{equation}
6A'^{2}+4A'B'-fe^{-8A}-6ke^{-2A}+\kappa^{2}\left(-\frac{1}{2}\phi'^{2}+V\right)=0\,,\quad\left(rr\right)\label{einsteinR}
\end{equation}

\begin{equation}
10A'^{2}+4A''-fe^{-8A}-6ke^{-2A}+\kappa^{2}\left(\frac{1}{2}\phi'^{2}+V\right)=0\,.\quad\left(\theta\theta\right)\label{einsteinPhi}
\end{equation}

To understand the back-reaction by the brane to the bulk, we also
need the brane-bulk matching conditions \cite{Burgess:2007vi,Burgess:2008yx,Bayntun:2009im},
which tells

\begin{equation}
\underset{r\rightarrow0}{\lim}\,\oint_{y_{b}}d\theta\,\sqrt{-g}\,\phi'=-\frac{\delta S_{brane}}{\delta\phi}\,,\label{scalarCon}
\end{equation}

\begin{equation}
\underset{r\rightarrow0}{\lim}\,\oint_{y_{b}}d\theta\,\sqrt{-g}\,F^{rM}=-\frac{\delta S_{brane}}{\delta A_{M}}\,,\label{fluxCon}
\end{equation}

\begin{equation}
\underset{r\rightarrow0}{\lim}\,\oint_{y_{b}}d\theta\,\frac{1}{2\kappa^{2}}\sqrt{-g}\left(K^{ij}-Kg^{ij}\right)-\left(\mathrm{flat}\right)=-\frac{\delta S_{brane}}{\delta g_{ij}}\,.\label{matchingGeneral}
\end{equation}

\noindent Here the extrinsic curvature $K_{ij}$ of fixed-$r$ surface
is given by $K_{ij}=\frac{1}{2}\partial_{r}g_{ij}$, with $i,j$ label
all coordinates except $r$. The ``flat'' is the same result substituting
$B=\ln r$ and $\underset{r\rightarrow0}{\lim}\,A'=0$ with $A\left(0\right)$
unchanged i.e. with a flat metric continuous to metric outside the
brane. The integration is along a small circle around the brane. Eq.
(\ref{fluxCon}) simply relates $\Phi_{b}$ to $F_{0}$ \cite{Burgess:2011mt,Burgess:2011va},
and for simplicity we will use $f$ to perform calculations. On the
other hand, Eq. (\ref{matchingGeneral}) becomes

\begin{equation}
-\underset{r\rightarrow0}{\lim}\,\frac{2\pi}{\kappa^{2}}e^{4A}\left(e^{B}\left(3A'+B'\right)-1\right)=T_{b}\,,\quad\left(\mu\nu\right)\label{matching1}
\end{equation}

\begin{equation}
\underset{r\rightarrow0}{\lim}\,\frac{2\pi}{\kappa^{2}}e^{B+4A}A'=-\frac{1}{2\sqrt{-\bar{g}}}\frac{\partial}{\partial g_{\theta\theta}}\left(\sqrt{\bar{g}}T_{b}\right)=U_{b}\,.\quad\left(\theta\theta\right)\label{matching2}
\end{equation}

\noindent From these matching conditions and the field equations,
one can derive a constraint \cite{Navarro:2004di,Burgess:2007vi,Burgess:2008yx,Bayntun:2009im}
in which $U_{b}$ can be fully determined by $T_{b}$ and $\frac{\partial T_{b}}{\partial\phi}$.
It reads

\begin{equation}
\frac{\kappa^{2}U_{b}}{2\pi}=\frac{1}{3}\left(e^{4A}-\frac{\kappa^{2}T_{b}}{2\pi}\pm\sqrt{\left(e^{4A}-\frac{\kappa^{2}T_{b}}{2\pi}\right)^{2}-\frac{3}{4}\left(\frac{\kappa^{2}}{2\pi}\frac{\partial T_{b}}{\partial\phi}\right)^{2}}\right)\,,\label{eq:braneconstraint}
\end{equation}

\noindent where the sign is chosen such that $U_{b}\rightarrow0$
when $\frac{\partial T_{b}}{\partial\phi}\rightarrow0$.

\subsubsection*{Graviton Modes}

After solving the background metric, we add linear perturbation to
solve for 4D graviton modes $h_{\mu\nu}$. The metric becomes

\begin{equation}
ds^{2}=e^{2A\left(r\right)}\left(\bar{g}_{\mu\nu}+h_{\mu\nu}\right)dx^{\mu}dx^{\nu}+dr^{2}+e^{2B\left(r\right)}d\theta^{2}\,.\label{perturbedMetric}
\end{equation}

By symmetry we can separate the variables as $h_{\mu\nu}\left(x,y\right)=\bar{h}_{\mu\nu}\left(x\right)\psi\left(y\right)$,
where in the transverse and traceless gauge it satisfies

\begin{equation}
\bar{h}_{\mu}^{\mu}=\bar{\nabla}^{\mu}\bar{h}_{\mu\nu}=0\,,
\end{equation}

\noindent where the covariant derivative is with respect to $\bar{g}$.
To study the spectrum of graviton modes, we would like $\left(\bar{\square}-\lambda\right)\bar{h}_{\mu\nu}=0$,
where $\bar{\square}$ is the 4D Laplacian with respect to $\bar{g}$
and $\lambda$ is the eigenvalue. The Pauli-Fierz mass of graviton
\cite{Buchbinder:1999ar,Deser:2001wx} is given by $m^{2}=\lambda-2k$.
By linearizing the field equations, $\psi$ satisfies \cite{Csaki:2000fc,Bachas:2011xa}

\begin{equation}
-e^{2A}\left(\psi''+\left(B'+4A'\right)\psi'+e^{-2B}\frac{\partial^{2}\psi}{\partial\theta^{2}}\right)=m^{2}\psi\,,\label{gravitonMode}
\end{equation}

\noindent where the dependence on background fields is hidden in the
warp factors. Now we can further decompose $\psi\left(y\right)=\bar{\psi}\left(r\right)e^{in\theta}$,
where $n$ is like a winding number. Note that in terms of $\Psi=e^{2A}\psi$,
Eq. (\ref{gravitonMode}) can be rewritten into a Schrodinger-like
equation:

\begin{equation}
\left(-\tilde{\square}+V\left(y\right)\right)\Psi\left(y\right)=m^{2}\Psi\left(y\right)\,,\quad V\left(y\right)=e^{-2A}\tilde{\square}e^{2A}\,,
\end{equation}

\noindent where the Laplacian $\tilde{\square}$ is with respect to
the inversely warped internal metric $\tilde{g}=\mathrm{diag}\,e^{-2A}\left(1,e^{2B}\right)$.
$\Psi$ also represents the amplitude of gravity in the extra dimensions.
Therefore, we usually refer $\Psi$ and even $\psi$ as wavefunctions,
but they are not quantum-mechanical wavefunctions.

When we say the graviton mode is localized, it means that $\Psi$
peaks at $r=0$ only and decays when it goes far from the brane. Its
norm is given by

\begin{equation}
\left\Vert \Psi\right\Vert ^{2}=\int dr\,d\theta\,e^{B+2A}\left|\psi\right|^{2}\,.
\end{equation}

\noindent To have physical graviton modes, the normalizability is
a required boundary condition at infinity. It is non-trivial since
we are integrating infinitely large proper radius $r$. There is also
a matching condition at the brane position.

Here is its derivation. Using Eq. (\ref{perturbedMetric}), we have

\begin{equation}
K_{\mu\nu}=A'e^{2A}\left(\bar{g}_{\mu\nu}+h_{\mu\nu}\right)+\frac{1}{2}e^{2A}h'_{\mu\nu}\,,
\end{equation}

\noindent while $K_{\theta\theta}$, $K$ and $\sqrt{-g}$ remain
unchanged by tracelessness of $h$. Therefore the $\theta\theta$-component
of Eq. (\ref{matchingGeneral}) is unchanged and the $\mu\nu$-component
becomes

\begin{equation}
-\underset{r\rightarrow0}{\lim}\,\frac{2\pi}{\kappa^{2}}e^{6A}\left[\left(e^{B}\left(3A'+B'\right)-1\right)\left(\bar{g}_{\mu\nu}+h_{\mu\nu}\right)-\frac{1}{2}\left(e^{B}-r\right)h'_{\mu\nu}\right]=e^{2A}T_{b}\left(\bar{g}_{\mu\nu}+h_{\mu\nu}\right)\,.
\end{equation}

\noindent The new matching condition for graviton modes is

\begin{equation}
\underset{r\rightarrow0}{\lim}\,e^{6A}\left(e^{B}-r\right)\psi'=0\,.\label{gravitonMatching}
\end{equation}

\noindent Note that by definition of branes $e^{B}-r\rightarrow0$
when $r\rightarrow0$, so a wavefunction with finite $\psi'\left(0\right)$
automatically satisfies the condition, but it remains interesting
to study the case that $\psi'\left(0\right)$ diverges.

In general the model cannot be solved analytically. To have a sense
on what kinds of geometry the solutions describe, below we consider
two exactly solvable cases. Both are with a bulk cosmological constant
$V=\frac{\Lambda}{\kappa^{2}}$ and without scalar fields i.e. $c=1$
and $U_{b}=0$, but one is with $k,f,B$ turned on only and another
is with $A,B$ turned on only. We will solve for their background
geometries and (seemingly valid) massless localized graviton modes.

\subsection{Exactly Solvable Models\label{subsec:twoeg}}

Here we study two examples to demonstrate the features of the above
setup.

\subsubsection*{With $\Lambda,k,f,B$}

Here we set $A$ to be constant, then Eq. (\ref{einsteinR}) and (\ref{einsteinPhi})
simply mean $\Lambda=fe^{-8A}+6ke^{-2A}$. Requiring $\underset{r\rightarrow0}{\lim}\,e^{B}=0$,
Eq. (\ref{einstein4d}) becomes

\begin{equation}
-9ke^{-2A}+2\Lambda+B'^{2}+B''=0\Rightarrow B\left(r\right)=\begin{cases}
\ln\left(\sin\left(\sqrt{2\Lambda-9ke^{-2A}}r\right)\right)+C_{1} & \Lambda>9ke^{-2A}/2\\
\ln r+C_{1} & \Lambda=9ke^{-2A}/2\\
\ln\left(\sinh\left(-\sqrt{-2\Lambda+9ke^{-2A}}r\right)\right)+C_{1} & \Lambda<9ke^{-2A}/2
\end{cases}\,,
\end{equation}

\noindent where $C_{1}$ is an integration constant. Since we are
studying noncompact extra dimensions, we only accept $\Lambda\leq\frac{9ke^{-2A}}{2}$.
The matching conditions reproduce the constraint on $B\left(0\right)$
in both cases and require

\begin{equation}
T_{b}=\begin{cases}
-\frac{2\pi}{\kappa^{2}}\left(e^{C_{1}}-1\right) & \Lambda=9ke^{-2A}/2\\
-\frac{2\pi}{\kappa^{2}}\left(\sqrt{-2\Lambda+9ke^{-2A}}e^{C_{1}}-1\right) & \Lambda<9ke^{-2A}/2
\end{cases}\,.
\end{equation}

\noindent Therefore this geometry is supported for all $T_{b}<\frac{2\pi}{\kappa^{2}}$
given the tuning of Maxwell field.

Now we solve for massless graviton modes. We focus on $n=0$. For
$\Lambda=\frac{9ke^{-2A}}{2}$,

\begin{equation}
\psi''+\frac{1}{r}\psi'=0\Rightarrow\psi\left(r\right)=C_{2}\ln r+C_{3}\,,
\end{equation}

\noindent which is clearly not localized and not normalizable for
all $C_{2}$ and $C_{3}$. For $\Lambda<\frac{9ke^{-2A}}{2}$,

\begin{equation}
\psi''+\sqrt{-2\Lambda+9ke^{-2A}}\coth\left(\sqrt{-2\Lambda+9ke^{-2A}}r\right)\psi'=0\Rightarrow\psi\left(r\right)=C_{2}\ln\left(\tanh\left(\sqrt{\frac{-2\Lambda+9ke^{-2A}}{4}}r\right)\right)+C_{3}\,,
\end{equation}

\noindent which is localized at $r=0$ and is normalizable if and
only if $C_{3}=0$:

\begin{align}
\left\Vert \Psi\right\Vert ^{2} & =2\pi\left|C_{2}\right|^{2}\int_{0}^{\infty}dr\,\sinh\left(\sqrt{-2\Lambda+9ke^{-2A}}r\right)\ln\left(\tanh\left(\sqrt{\frac{-2\Lambda+9ke^{-2A}}{4}}r\right)\right)^{2}\nonumber \\
 & =10.3354\frac{\left|C_{2}\right|^{2}}{\sqrt{-2\Lambda+9ke^{-2A}}}<\infty\,.
\end{align}

\noindent However, the matching condition reads

\begin{equation}
\underset{r\rightarrow0}{\lim}\,\left(e^{B}-r\right)\psi'=C_{2}\left(\sqrt{-2\Lambda+9ke^{-2A}}e^{C_{1}}-1\right)=-\frac{\kappa^{2}}{2\pi}C_{2}T_{b}=0\,.
\end{equation}

\noindent Since $T_{b}\neq0$, we also have $C_{2}=0$ and there does
not exist any massless graviton mode which satisfies all boundary
conditions.

\subsubsection*{With $\Lambda,A,B$}

Here we set $f=k=0$ and non-constant $A$. Solving Eq. (\ref{einsteinR})
and (\ref{einsteinPhi}) and imposing $\underset{r\rightarrow0}{\lim}\,e^{B}=0$
yields

\begin{gather}
A\left(r\right)=\begin{cases}
\frac{2}{5}\ln r+C_{1} & \Lambda=0\\
\frac{2}{5}\ln\left(\cosh\left(\sqrt{\frac{-5\Lambda}{8}}r\right)\right)+C_{1} & \Lambda<0
\end{cases}\,,\\
B\left(r\right)=\begin{cases}
-\frac{3}{5}\ln r+C_{2} & \Lambda=0\\
-\frac{3}{5}\ln\left(\cosh\sqrt{\frac{-5\Lambda}{8}}r\right)+\ln\left(\sinh\sqrt{\frac{-5\Lambda}{8}}r\right)+C_{2} & \Lambda<0
\end{cases}\,.
\end{gather}

\noindent These solutions automatically also satisfy Eq. (\ref{einstein4d}).
With the same reason as above we do not consider $\Lambda>0$. However
for $\Lambda=0$, Eq. (\ref{matching2}) becomes $e^{4C_{1}+C_{2}}=0$,
which cannot be satisfied. Therefore this geometry is not supported
due to matching conditions. Even if it is supported, the wavefunction
is not normalizable under this geometry as above.

For $\Lambda<0$, the solution is usually known as the AdS soliton
\cite{Horowitz:1998ha}. The matching conditions lead to

\begin{equation}
T_{b}=-\frac{\pi e^{4C_{1}}\left(\sqrt{-10\Lambda}e^{C_{2}}-4\right)}{2\kappa^{2}}\,.
\end{equation}

\noindent Therefore this geometry is valid for all $\Lambda\leq0$
and $T_{b}$. Now the massless $n=0$ graviton mode satisfies

\begin{equation}
\psi''+\sqrt{\frac{-5\Lambda}{2}}\coth\left(\sqrt{\frac{-5\Lambda}{2}}r\right)\psi'=0\Rightarrow\psi\left(r\right)=C_{3}\ln\left(\tanh\left(\sqrt{\frac{-5\Lambda}{8}}r\right)\right)+C_{4}\,.
\end{equation}

\noindent This is a more extreme example than above. Even when the
warping amplifies the localized wavefunction, it can still be normalized
when $C_{4}=0$:

\begin{align}
\left\Vert \Psi\right\Vert ^{2} & =2\pi\left|C_{3}\right|^{2}\int_{0}^{\infty}dr\,\sinh\left(\sqrt{\frac{-5\Lambda}{8}}r\right)\cosh^{1/5}\left(\sqrt{\frac{-5\Lambda}{8}}r\right)\ln\left(\tanh\left(\sqrt{\frac{-5\Lambda}{8}}r\right)\right)^{2}\nonumber \\
 & =2.80403\frac{\left|C_{3}\right|^{2}}{\sqrt{-\Lambda}}<\infty\,.
\end{align}

\noindent However, the matching condition reads

\begin{equation}
\underset{r\rightarrow0}{\lim}\,e^{6A}\left(e^{B}-r\right)\psi'=\frac{C_{3}e^{6C_{1}}\left(\sqrt{-10\Lambda}e^{C_{2}}-4\right)}{4}=-\frac{\kappa^{2}}{2\pi}e^{2C_{1}}C_{3}T_{b}=0\,.
\end{equation}

\noindent Again, it forces $C_{3}=0$ and there is not any massless
graviton modes.

As a remark, there is another interesting solution to this background,
which is

\begin{equation}
A=C_{1}-\sqrt{\frac{-\Lambda}{10}}r\,,\quad B=C_{2}-\sqrt{\frac{-\Lambda}{10}}r\,.\label{eq:RS}
\end{equation}

\noindent This is the only solution with a decaying warp factor $e^{2A}$,
hence the constant wavefunction can be localized and normalized. On
the other hand, it does not satisfy $\underset{r\rightarrow0}{\lim}\,e^{B}=0$.
Notice that the ordinary type-II Randall-Sundrum (RS-II) model \cite{Randall:1999vf}
is equivalent to the constant wavefunction in our formulation. This
is why we cannot construct such model with two noncompact extra dimensions.
However, we will see that this solution has another related implication
when combined with compactification.

\subsection{Necessary Conditions for Existence of Massless Localized Graviton\label{subsec:Necessary-Conditions}}

We already see that localized $\psi$ is rejected by the matching
condition in the above examples. Here we consider back the general
setup in Section \ref{subsec:General-Setup} and prove the necessary
conditions in order to localize a massless graviton.

Consider the wavefunction of a massless graviton. By separation of
variables, $\psi\left(y\right)=\bar{\psi}\left(r\right)e^{in\theta}$
where $n$ is integer. We have that $\bar{\psi}\left(0\right)e^{in\theta}$
are equal for all $\theta$, which means $\bar{\psi}\left(0\right)=0$
and the graviton mode cannot be localized when $n\neq0$. Therefore
we only consider $n=0$. Eq. (\ref{gravitonMode}) implies

\begin{equation}
\psi''+\left(B'+4A'\right)\psi'=0\Rightarrow\psi'=Ce^{-B-4A}\,,
\end{equation}

\noindent where $C$ is an integration constant. Using the fact that
$A\left(0\right)$ is finite in order to have physical 4D induced
metric, the matching condition Eq. (\ref{gravitonMatching}) reads,

\begin{equation}
\underset{r\rightarrow0}{\lim}\,C\left(1-\frac{r}{e^{B}}\right)=0\,.
\end{equation}

There is a subtlety that there are two ways to understand the matching
condition. The first one is to note that since the matching condition
is obtained by integrating a small circle around brane position, when
we say $r\rightarrow0$ we actually mean substituting some small $r=\epsilon$.
Since in curved background $\frac{\epsilon}{e^{B\left(\epsilon\right)}}\neq1$,
the matching condition simply implies $C=0$ i.e. constant wavefunction.
This is actually an intuitive statement since there is no brane sources
to drive the massless graviton modes.

The second one is to directly take the $r\rightarrow0$ limit. We
then need to be careful what the ratio $\frac{r}{e^{B}}$ contributes.
Note that this ratio captures the defect angle at brane position.
To be precise, the defect angle $\delta=2\pi\left(1-\alpha\right)$
is given by

\begin{equation}
\alpha=\underset{r\rightarrow0}{\lim}\,\frac{e^{B}}{r}\,.
\end{equation}

Therefore the matching condition implies that $C=0$ unless there
is no defect angle, which is not true for typical brane sources. We
thus conclude that $\psi$ must be a constant. If the warp factor
$e^{B+2A}$ does not vanish at infinity, the constant wavefunction
is not normalizable and there is no massless graviton modes. Therefore
we state: 
\begin{quote}
\begin{center}
Massless localized graviton exists only when $\int dr\,d\theta\,e^{B+2A}$
is finite. 
\par\end{center}
\end{quote}

\subsection{No-Go Theorems\label{subsec:Behaviour-of-Warp}}

\subsubsection*{Two noncompact extra dimensions}

The constant wavefunction can only be localized and normalized by
the warp factor $e^{2A}$. Through this, we show that within our model,
the constant wavefunction is not a valid solution.

First, we assume that $e^{A}$ is decreasing to small values at large
$r$, hence $A$ is decreasing at large $r$. By combining Eq. (\ref{einsteinR})
and (\ref{einsteinPhi}), for sufficiently large $r>r_{1}$ we have

\begin{equation}
A''+A'^{2}=A'B'-\frac{\kappa^{2}}{4}\phi'^{2}<0\,,\label{directProof}
\end{equation}

\noindent since $B'>0$ at large $r$. It means that $\left(e^{A}\right)''=e^{A}\left(A''+A'^{2}\right)<0$
for $r>r_{1}$. Then for all $y>x>r_{1}$, if $x,y$ are within the
domain of $A$, by Mean Value Theorem there exists $\xi\in\left(x,y\right)$
such that

\begin{equation}
\frac{e^{A\left(y\right)}-e^{A\left(x\right)}}{y-x}=\left(e^{A}\right)'\left(\xi\right)\leq\left(e^{A}\right)'\left(x\right)\Rightarrow e^{A\left(y\right)}\leq e^{A\left(x\right)}+\left(e^{A}\right)'\left(x\right)\left(y-x\right)\,.\label{MVT-2}
\end{equation}

\noindent Since $\left(e^{A}\right)'\left(x\right)<0$, by choosing
sufficiently large $y$ such that $e^{A\left(x\right)}+\left(e^{A}\right)'\left(x\right)\left(y-x\right)<0$,
Eq. (\ref{MVT-2}) means that $e^{A}$ must vanish at some finite
$r<y$, but not infinity. This implies that there is no solution at
infinity, which violates the assumption of noncompact extra dimensions.
By contradiction, it means that $e^{A}$ does not vanish at infinity
in noncompact extra dimensions, thus cannot localize or normalize
the constant wavefunction.

This finishes the proof of the no-go theorem on localized gravity
around codimension-2 branes in two noncompact extra dimensions. However
as we will see, this proof makes use of unique features in codimension-2
models, and cannot be generalized to higher codimensions.

\subsubsection*{One Compact and One Noncompact Extra Dimensions}

So far we have studied the possibility of localized gravity with extra
dimensions being noncompact in two directions i.e. $B\rightarrow+\infty$.
It still remains interesting to explore the case when $B$ keeps finite,
such that at long distances the spacetime looks like with one noncompact
extra dimension, but with another one compactified dimension. For
example, we can attach a small circle at each point in the Randall-Sundrum
model, such that the extra dimensions have topology $\mathbb{R}\times S^{1}$,
and the geometry looks like a long thin tube. In such cases, at long
distances we cannot really tell whether the brane is codimension-1
or codimension-2.

If the brane is still codimension-1, that means the brane wraps around
the hidden circle. Such geometry actually can be easily obtained,
such as in Eq. (\ref{eq:RS}). We identify $r$ as the height of the
cylinder and $e^{B}$ as the compactification radius. We further replace
$r$ by $\left|r\right|$ in those equations. The compact extra dimension
becomes smaller and smaller as $r$ increases, so by fixing suitable
$C_{2}$ we can keep the compact extra dimension very small for all
$r$. Now if we put a wrapped 4-brane at $r=0$, with appropriate
matching conditions the constant massless graviton wavefunction is
localized around the circle $r=0$ by the warp factor $A$. It is
equivalent to the ordinary RS-II model \cite{Randall:1999vf} but
in two extra dimensions with one being compactified.

If the brane is actually codimension-2, at $r=0$ the hidden dimension
must shrink to a point, in order to maintain the symmetry of the system.
This means that the topology of the bulk is changed, and it is non-trivial
whether the long distance physics remain the same. Below we provide
a non-rigorous argument to show that it is not possible to both achieve
such geometry and localize gravity.

Again we first assume that localized gravity in such geometry is possible.
We start with the behavior of $A$ and $B$. In this case we still
have $\underset{r\rightarrow0}{\lim}\,e^{B}=0$ i.e. $B\left(0\right)\rightarrow-\infty$.
Since at $r=0$ we have $\left(e^{B}\right)'>0$, it means $B'\left(0\right)\rightarrow+\infty$.
To localize gravity at brane position, we want $A'\left(0\right)$
to be negative. Then by Eq. (\ref{directProof}), we get $A''\left(0\right)\rightarrow-\infty$.
In particular it means that $A'\left(0\right)$ needs to be finite.
Eq. (\ref{matching2}) then tells $U_{b}=0$, which implies $\frac{\partial T_{b}}{\partial\phi}=0$
by Eq. (\ref{eq:braneconstraint}).

An increasing $B$ has already been forbidden in last subsection.
Therefore, $B$ must turn decreasing at some finite $r$. After $B$
starts decreasing, it must fall not too fast such that $e^{B}$ does
not vanish at some non-zero and finite $r$. This can be judged by
combining Eq. (\ref{einstein4d}) to (\ref{einsteinPhi}), which gives

\begin{equation}
e^{-B}\left(e^{B}\right)''=B''+B'^{2}=-4A'B'-\frac{3}{2}fe^{-8A}-\frac{\kappa^{2}}{2}V\,.
\end{equation}

Now we consider the behavior of the potential $V$. From Eq. (\ref{einsteinR}),
we observe that

\begin{equation}
V\left(r=0\right)=\left.\frac{1}{\kappa^{2}}\left(-6A'^{2}-4A'B'+fe^{-8A}+6ke^{-2A}\right)+\frac{1}{2}\phi'^{2}\right|_{r=0}\rightarrow+\infty\,.
\end{equation}

\noindent If $V$ keeps positive for all $r$, after $B$ starts decreasing
at some finite $r$, we have $\left(e^{B}\right)''<0$ and $e^{B}$
vanishes at another finite $r$ by the argument in last subsection.
Therefore to meet the above requirement $V$ must turn negative at
some finite $r=r_{2}$. It means that there is always a region of
small $r$ between $0$ and $r_{2}$ that has negative $V'$ with
large magnitude. In particular, there must be a point where $V'$
diverges to minus infinity, since $V$ starts at positive infinity.

We then turn to the field equation for the scalar field i.e. Eq. (\ref{scalarFieldEqn}).
Multiplying it with $\phi'$, we get

\begin{equation}
\left(\frac{1}{2}\phi'^{2}\right)'+\left(B'+4A'\right)\phi'^{2}=V'+\frac{1}{4}c'\frac{f}{c}e^{-8A}\,.\label{eq:scalarFieldFinal}
\end{equation}

\noindent At small $r$ and smaller than $r_{2}$, $B'+4A'$ is positive
while the $f$ term is suppressed by large $e^{A}$. Therefore in
that region the kinetic energy $\frac{1}{2}\phi'^{2}$ decreases rapidly,
and even infinitely fast at the point where $V'$ diverges. However,
since $\frac{1}{2}\phi'^{2}$ is always positive, it means that $\phi'$
must start at very large value at $r=0$. By integrating Eq. (\ref{eq:scalarFieldFinal}),
we see that $\phi'$ actually starts at infinity, which implies $\frac{\partial T_{b}}{\partial\phi}\neq0$
by Eq. (\ref{scalarCon}). Note that we now have a contradiction,
so we can conclude that we cannot both obtain a long thin tube geometry
and localize gravity around a codimension-2 brane in that geometry.
Notice that here we have used many properties specific to two extra
dimensions, and the proof cannot be generalized to more than one compact
dimensions.

In conclusion, within our setup, whenever we see a codimension-1 model
but with a hidden compact dimension, we can claim that the brane must
also wrap that dimension i.e. it must be also codimension-1 but not
codimension-2. The only difference between two scenarios is the topology
of the extra dimensions. This result reveals the typical behavior
that the topological details of compactified extra dimensions can
affect the low-energy physics, even when we cannot observe those dimensions.

\section{Branes with Higher Codimensions\label{sec:Branes-with-Higher}}

In this section, we investigate the no-go theorem in codimensions
higher than two.

\subsection{Klebanov-Strassler Throat}

After the above proof, it is natural to consider whether localization
of gravity in more than two noncompact extra dimensions is still possible.
In particular, such possibility in string theory solution is important
to string phenomenology. Here as an example, we consider the Klebanov-Strassler
throat \cite{Klebanov:2000hb}. Similarly to above, we focus on the
UV region, but not the deformed IR region of the throat. Therefore
for simplicity, we only consider the undeformed version of the throat
i.e. the geometry in \cite{Klebanov:2000nc}, and similar but more
complicated arguments should hold for that in \cite{Klebanov:2000hb}.
The metric is given by

\begin{equation}
ds_{10}^{2}=e^{2a\left(u\right)-5q\left(u\right)}\eta_{\mu\nu}dx^{\mu}dx^{\nu}+e^{-5q\left(u\right)}du^{2}+e^{3q\left(u\right)}ds_{T^{1,1}}^{2}\,.
\end{equation}

\noindent Here the functions $a$ and $q$ are analog to $A$ and
$B$ in our codimension-2 context, $u$ is the radial coordinate of
the throat and $T^{1,1}=\left(SU\left(2\right)\times SU\left(2\right)\right)/U\left(1\right)$
is the base of the conifold with metric

\begin{equation}
ds_{T^{1,1}}^{2}=\frac{1}{9}\left(d\psi+\sum_{i=1}^{2}\cos\theta_{i}\,d\phi_{i}\right)^{2}+\frac{1}{6}\sum_{i=1}^{2}\left(d\theta_{i}^{2}+\sin^{2}\theta_{i}\,d\phi_{i}^{2}\right)\,.
\end{equation}

The norm of constant wavefunction is given by

\begin{equation}
\left\Vert \Psi\right\Vert ^{2}=\int d^{6}x\,\sqrt{e^{-5q}\left(e^{3q}\right)^{5}}e^{2a-5q}\propto\int du\,e^{2a}\,.
\end{equation}

\noindent Therefore to study the normalizability, we study whether
$e^{2a}$ can decay to zero at infinity. By solving the type IIB supergravity
equations, we have

\begin{equation}
a\left(u\right)=A_{0}+q\left(u\right)+\frac{1}{P}T\left(u\right)\,,
\end{equation}

\noindent where the constant $P>0$ and $T\left(u\right)$ are related
to the R-R 3-form field strength and NS-NS 2-form potential respectively,
$A_{0}$ is an integration constant. There is another solution with
$a=A_{0}+u$, which is clearly not normalizable. Define $Y=e^{6q}$
and $K=4+PT$ ($K$ is actually related to the self-dual 5-form field
strength), the field equations give 
\begin{equation}
K'=P^{2}Y^{-2/3},\quad\frac{dY}{dK}=\frac{1}{P^{2}}\left(4Y-K\right)\,,
\end{equation}

\noindent where here ``prime'' represents derivative with respect
to $u$. There is a general solution for $Y$:

\begin{equation}
Y=a_{0}e^{4K/P^{2}}+\frac{K}{4}+\frac{P^{2}}{16}\,,
\end{equation}

\noindent where $a_{0}$ is an integration constant. Note that $K$
is always increasing, so as $T$. If $a_{0}\geq0$, $Y$ is also increasing,
thus $a$ is increasing and $e^{2a}$ cannot decay to zero at infinity.
Now we focus on the case with $a_{0}<0$ and $Y$ is not always increasing.
It turns out that this case violates the assumption of noncompact
extra dimensions.

The proof is as follows. Let $u=u_{1}$ be a point such that $Y$
is strictly decreasing. For all $u>u_{1}$ and within domain of $Y$,

\begin{equation}
Y'=\frac{dY}{dK}K'=Y^{-2/3}\left(4a_{0}e^{4K\left(u\right)/P^{2}}+\frac{P^{2}}{4}\right)\leq Y^{-2/3}\left(4a_{0}e^{4K\left(u_{1}\right)/P^{2}}+\frac{P^{2}}{4}\right)\leq0\,,
\end{equation}

\noindent and

\begin{equation}
K''=\frac{d\left(K'\right)}{dY}Y'=-\frac{2}{3}P^{2}Y^{-5/3}Y'\geq0\,.
\end{equation}

\noindent Therefore $K'\geq0$ and $K''\geq0$ for all such $u$.
Now for all $y>x>u_{1}$, if $x,y$ are within the domain of $A$,
by Mean Value Theorem there exists $\xi\in\left(x,y\right)$ such
that

\begin{equation}
\frac{K\left(y\right)-K\left(x\right)}{y-x}=K'\left(\xi\right)\geq K'\left(x\right)\Rightarrow K\left(y\right)\geq K\left(x\right)+K'\left(x\right)\left(y-x\right)\,.\label{MVT-1}
\end{equation}

\noindent Let $Y=0$ at finite $K=K_{1}$. By choosing sufficiently
large $y$, Eq. (\ref{MVT-1}) guarantees that $K$ must reach $K_{1}$
at some finite $u=u_{2}$. In other words, $Y$ decreases to zero
at $u_{2}$ and there is no solution for $u>u_{2}$. Furthermore,
numerical integration shows that the proper radius $\int^{u_{2}}du\,e^{-5q/2}=\int^{u_{2}}du\,Y^{-5/12}$
is finite. Therefore we conclude that $Y$ can only be increasing
in noncompact extra dimensions, and constant wavefunction is not normalizable.
Therefore, the only way to localize gravity with the Klebanov-Strassler
throat is to make a UV cutoff of the throat, as how the throat is
attached to a compact manifold in the context of compactification.

\subsection{General Setup}

Before we prove the no-go theorem in general, let us first rewrite
some of the general setup in Section \ref{subsec:General-Setup} for
codimension-$d$ branes where $d\geq3$. We let the bulk Lagrangian
be

\begin{equation}
S_{bulk}=-\int d^{4}x\,d^{d}y\,\sqrt{-g}\left(\frac{1}{2\kappa^{2}}R+\frac{1}{2}\partial^{M}\phi\partial_{M}\phi+\sum_{p}\frac{1}{2p}c_{p}\left(\phi\right)F_{p}^{2}+V\left(\phi\right)\right)+S_{GH}\,,
\end{equation}

\noindent where $2\leq p\leq d$ and $F_{p}=dA_{p-1}$ is a $p$-form
field strength. We use the notation $F_{p}^{2}=F_{M_{1}M_{2}...M_{p}}F^{M_{1}M_{2}...M_{p}}$.
We first study the case where all extra dimensions are noncompact.
The metric becomes

\begin{equation}
ds^{2}=g_{MN}dx^{M}dx^{N}=e^{2A\left(r\right)}\bar{g}_{\mu\nu}dx^{\mu}dx^{\nu}+dr^{2}+e^{2B\left(r\right)}\hat{g}_{ab}\left(\theta\right)d\theta^{a}d\theta^{b}\,,
\end{equation}

\noindent where $1\leq a,b\leq d-1$ and $\hat{g}$ is the metric
for internal angular coordinates. By symmetry $A_{p-1}$ depends on
$r$ only, and the non-zero components of $F_{p}$ are

\begin{equation}
F_{ra_{1}a_{2}...}\,,\quad F_{\mu\nu\rho\sigma ra_{1}a_{2}...}\propto\epsilon_{\mu\nu\rho\sigma}\,.\label{eq:possibleF}
\end{equation}

\noindent Here $a_{1},a_{2},...$ are angular coordinates and $\epsilon_{\mu\nu\rho\sigma}$
is the Levi-Civita tensor in brane directions. The latter exists only
when $p\geq5$. Such field strengths do appear in, for example, type
IIB string theory.

Let us now consider the brane-bulk matching conditions. In higher
codimensions, it is natural to generalize Eq. (\ref{matching1}) to

\begin{equation}
-\underset{r\rightarrow0}{\lim}\,\frac{2\pi^{d/2}}{\Gamma\left(d/2\right)\kappa^{2}}e^{\left(d-1\right)B+4A}\left(3A'+\left(d-1\right)B'\right)=T_{b}\,,\quad\left(\mu\nu\right)
\end{equation}

\noindent and the boundary condition for $\psi'$ is

\begin{equation}
\underset{r\rightarrow0}{\lim}\,e^{6A}\left(e^{\left(d-1\right)B}-r^{d-1}\right)\psi'=0\,.
\end{equation}

The graviton mode equation is \cite{Csaki:2000fc,Bachas:2011xa}

\begin{equation}
-e^{2A}\left(\psi''+\left(\left(d-1\right)B'+\left(d+2\right)A'\right)\psi'+e^{-2B}\hat{\square}\psi\right)=m^{2}\psi\,,
\end{equation}

\noindent where $\hat{\square}$ is the Laplacian with respect to
$\hat{g}$. As in Section \ref{subsec:Necessary-Conditions}, massless
localized $\psi$ only depends on $r$, then we have $\psi'=Ce^{-\left(d-1\right)B-\left(d+2\right)A}$
where $C$ is the integration constant, the boundary condition becomes

\begin{equation}
\underset{r\rightarrow0}{\lim}\,C\left(1-\left(\frac{r}{e^{B}}\right)^{d-1}\right)=0\,,
\end{equation}

\noindent which is satisfied in curved background only if $C=0$ i.e.
$\psi$ is constant. Therefore the statement in Section \ref{subsec:Necessary-Conditions}
also holds in higher codimensions. The norm of $\psi$ is now defined
as \cite{Bachas:2011xa}

\begin{equation}
\left\Vert \Psi\right\Vert ^{2}=\int dr\,\prod_{i}d\theta^{i}\,\sqrt{\hat{g}}e^{\left(d-1\right)B+2A}\left|\psi\right|^{2}\,.
\end{equation}

\noindent In order to have normalizable solutions, we thus require
$e^{\left(d-1\right)B+2A}$ to vanish at infinity.

We next study the field equations. Those for the Maxwell fields are

\begin{equation}
\nabla_{M}\left(c_{p}F^{MN_{1}N_{2}...N_{p-1}}\right)=0\,.
\end{equation}

\noindent These yield schematically,

\begin{equation}
c_{p}F_{p}^{2}=\frac{f_{1p}^{2}}{c_{p}}e^{-2\left(d-p\right)B-8A}+\frac{f_{2p}^{2}}{c_{p}}e^{-2\left(d-p+4\right)B}\,,
\end{equation}

\noindent where $f_{1p}$ and $f_{2p}$ are some finite functions
of the coordinates other than $r$. They are contributed by the former
and the latter in Eq. (\ref{eq:possibleF}) respectively. At large
$r$, the $f_{2p}$ term is suppressed by large $e^{B}$, so we can
safely ignore it for our purpose as long as $f_{1p}\neq0$. Below
we just assume $c_{p}F_{p}^{2}\propto c_{p}^{-1}e^{-2\left(d-p\right)B-8A}$.
Next we consider the $ab$-components of the Einstein's equation:

\begin{align}
 & g_{ab}\left(10A'^{2}+4\left(d-2\right)A'B'+4A''+\frac{\left(d-1\right)\left(d-2\right)}{2}B'^{2}+\left(d-2\right)B''-6ke^{-2A}\right)-\hat{R}_{ab}+\frac{1}{2}\hat{R}\hat{g}_{ab}\nonumber \\
= & \underset{p}{\sum}c_{p}\kappa^{2}\left(F_{aX_{1}X_{2}...X_{p-1}}\mathit{F_{b}}^{X_{1}X_{2}...X_{p-1}}-\frac{1}{2p}g_{ab}F_{p}^{2}\right)-\kappa^{2}g_{ab}\left(\frac{1}{2}\phi'^{2}+V\right)\,,\label{fieldEqnHigherD}
\end{align}

\noindent where $a,b$ label the angular coordinates and $X$ labels
all internal coordinates. $\hat{R}_{ab}$ and $\hat{R}$ are the Ricci
tensor and Ricci scalar with respect to $\hat{g}$ respectively. We
expect $\hat{R}$ to be a finite function of $\theta$. We then take
a partial trace i.e. contracting Eq. (\ref{fieldEqnHigherD}) with
$g^{ab}$ to reach

\begin{align}
 & \left(d-1\right)\left(10A'^{2}+4\left(d-2\right)A'B'+4A''+\frac{\left(d-1\right)\left(d-2\right)}{2}B'^{2}+\left(d-2\right)B''\right)\nonumber \\
= & \left(d-1\right)\left(6ke^{-2A}-\kappa^{2}\left(\frac{1}{2}\phi'^{2}+V\right)\right)+\sum_{p}c_{p}\kappa^{2}\left(1-\frac{d+1}{2p}\right)F_{p}^{2}-\frac{d-3}{2}\hat{R}e^{-2B}=-J\,,\label{tracedFieldEqnHigherD}
\end{align}

\noindent where we have used $F_{aX_{1}X_{2}...X_{p-1}}F^{aX_{1}X_{2}...X_{p-1}}=\left(1-\frac{1}{p}\right)F_{p}^{2}$.
We have also defined a quantity $J$ which is useful later. Similarly,
the $rr$-component is

\begin{align}
 & 6A'^{2}+4\left(d-1\right)A'B'+\frac{\left(d-1\right)\left(d-2\right)}{2}B'^{2}\nonumber \\
= & 6ke^{-2A}-\kappa^{2}\left(-\frac{1}{2}\phi'^{2}+V\right)+\sum_{p}c_{p}\kappa^{2}\frac{1}{2p}F_{p}^{2}-\frac{1}{2}\hat{R}e^{-2B}\,.\label{fieldEqnHigherDrr}
\end{align}

\subsection{No-Go Theorem}

Now we show that the constant graviton wavefunction is not a valid
solution, similarly to Section \ref{subsec:Behaviour-of-Warp}. First
we try to derive an equation analogous to Eq. (\ref{directProof}).
By combining Eq. (\ref{tracedFieldEqnHigherD}) and (\ref{fieldEqnHigherDrr}),
we have

\begin{equation}
\left(d-1\right)\left(4A'^{2}-4A'B'+4A''+\left(d-2\right)B''\right)=-\left(d-1\right)\kappa^{2}\phi'^{2}+\sum_{p}c_{p}\kappa^{2}\left(1-\frac{d}{p}\right)F_{p}^{2}+\hat{R}e^{-2B}\,.\label{eq:notDirectProof}
\end{equation}

\noindent The main obstruction of proving a similar no-go theorem
as in Section \ref{subsec:Behaviour-of-Warp} is the $B''$ term,
which does not appear for codimension-2 branes. Although it is not
typical, noncompact extra dimensions do not stop $B''$ from being
negative with large magnitude. The $\hat{R}$ term can also be positive.
Therefore, we will prove the no-go theorem in a different way with
some additional assumptions. The assumptions are
\begin{itemize}
\item There is always a form field background with $f_{1p}^{2}>0$ for some
$p$ and nonvanishing couplings. Mathematically, it means that $c_{p}$
is bounded from above, so $f_{1p}^{2}/c_{p}$ is bounded from below
by a positive number. 
\end{itemize}
or 
\begin{itemize}
\item The spacetime in brane directions is curved i.e. $k\neq0$. 
\end{itemize}
Such background is common in string theory (motivated) setups. We
also assume 
\begin{itemize}
\item Some of the terms are dominant in $J$ at large $r$. In other words,
$J$ does not oscillate between positive and negative values at large
$r$. It enables us to use arguments similar to the proof of Maldacena-Nunez
no-go theorem \cite{Maldacena:2000mw} on existence of dS compactifications.
\end{itemize}
Here is an outline of the proof: 
\begin{itemize}
\item We first assume that $e^{\left(d-1\right)B+2A}$ is decreasing to
small values at large $r$, hence $A$ is also decreasing at large
$r$. We then prove that for all backgrounds following the above assumptions,
$e^{\left(d-2\right)B/2+2A}$, and thus $e^{\left(d-1\right)B+2A}$
can decrease to zero only at finite $r$, but not infinity. As in
Section \ref{subsec:Behaviour-of-Warp}, it means that the constant
wavefunction cannot be localized or normalized. We divide the backgrounds
into several cases: 
\item The dominant term in $J$ at large $r$ is positive: We construct
upper bounds of $e^{\left(d-2\right)B/2+2A}$ by concavity to show
that it must vanish at some finite $r$ instead of infinity. 
\item The dominant term in $J$ at large $r$ is negative: The form of the
possible dominant terms and the assumptions give a differential inequality.
It again leads to upper bounds of $e^{\left(d-2\right)B/2+2A}$ showing
that it must vanish at some finite $r$ instead of infinity. 
\item There may be more than one terms that are as dominant as each other.
This only modifies the numerical coefficient of the asymptotic form
of $J$, which does not change the above arguments. 
\item Combining the above cases, we can conclude for all the above backgrounds,
the constant wavefunction is not a valid solution. 
\end{itemize}
Let us now fill in the details of the proof.
\begin{itemize}
\item The dominant term in $J$ at large $r$ is positive: 
\end{itemize}
We have $J>0$ at large $r$. Note that we can rewrite the left hand
side of Eq. (\ref{tracedFieldEqnHigherD}) into

\begin{align}
 & \left(d-1\right)\left(10A'^{2}+4\left(d-2\right)A'B'+4A''+\frac{\left(d-1\right)\left(d-2\right)}{2}B'^{2}+\left(d-2\right)B''\right)\nonumber \\
= & \left(d-1\right)\left(2A'^{2}+\frac{d-2}{2}B'^{2}+2\left(\frac{d-2}{2}B''+2A''+\left(\frac{d-2}{2}B'+2A'\right)^{2}\right)\right)\nonumber \\
= & \left(d-1\right)\left(2A'^{2}+\frac{d-2}{2}B'^{2}+2e^{-\left(d-2\right)B/2-2A}\left(e^{\left(d-2\right)B/2+2A}\right)''\right)
\end{align}

\noindent Therefore $\left(e^{\left(d-2\right)B/2+2A}\right)''$ is
negative at large $r$ and $e^{\left(d-2\right)B/2+2A}$ cannot vanish
at infinity, according to the arguments in Section \ref{subsec:Behaviour-of-Warp}.
\begin{itemize}
\item The dominant term in $J$ at large $r$ is negative: 
\end{itemize}
We first stick with the first assumption on form fields. From Eq.
(\ref{eq:notDirectProof}),

\begin{equation}
\left(d-1\right)\left(-4A'B'+4A''+\left(d-2\right)B''\right)\leq\hat{R}e^{-2B}\,.
\end{equation}

Now there is a negative term $\propto e^{-2\left(d-p\right)B-8A}$
in $J$. No matter whether that term is dominant in $J$, there must
exist a constant $v_{0}>0$ such that $-J-\hat{R}e^{-2B}\geq v_{0}e^{-2\left(d-p\right)B-8A}\geq v_{0}e^{-2\left(d-2\right)B-8A}$
at large $r$. Then from Eq. (\ref{tracedFieldEqnHigherD}), at large
$r$ we have

\begin{align}
 & \left(d-1\right)\left(10A'^{2}+4\left(d-1\right)A'B'+\frac{\left(d-1\right)\left(d-2\right)}{2}B'^{2}\right)\nonumber \\
= & -J-\left(d-1\right)\left(-4A'B'+4A''+\left(d-2\right)B''\right)\geq v_{0}e^{-2\left(d-2\right)B-8A}\,.\label{eq:Jbound}
\end{align}

In addition, since $e^{\left(d-1\right)B+2A}$ is decreasing at large
$r$, we have $\left(d-1\right)A'B'+2A'^{2}\geq0$ at large $r$.
Straightforward computation then leads to

\begin{align}
 & \frac{2\left(d-1\right)}{\min\left\{ 4,d-2\right\} }\left(\frac{d-2}{2}B'+2A'\right)^{2}\nonumber \\
= & \begin{cases}
8\frac{d-1}{d-2}A'^{2}+4\left(d-1\right)A'B'+\frac{\left(d-1\right)\left(d-2\right)}{2}B'^{2} & d\leq6\\
10A'^{2}+4\left(d-1\right)A'B'+\frac{\left(d-1\right)\left(d-2\right)^{2}}{8}B'^{2}+\left(d-6\right)\left(\left(d-1\right)A'B'+2A'^{2}\right) & d\geq6
\end{cases}\nonumber \\
\geq & 10A'^{2}+4\left(d-1\right)A'B'+\frac{\left(d-1\right)\left(d-2\right)}{2}B'^{2}\,.
\end{align}

\noindent We then reach

\begin{equation}
\left(\frac{d-2}{2}B'+2A'\right)^{2}\geq ve^{-4\left(\left(d-2\right)B/2+2A\right)}\,,\label{eq:DInequality}
\end{equation}

\noindent where $v=v_{0}\frac{\min\left\{ 4,d-2\right\} }{2\left(d-1\right)^{2}}$
is another positive constant. Since $e^{\left(d-2\right)B/2+2A}$
is decreasing at large $r$, the above can be simplified to

\begin{equation}
\left(e^{\left(d-2\right)B+4A}\right)'\leq-2\sqrt{v}\,.
\end{equation}

\noindent Therefore by Mean Value Theorem, $e^{\left(d-2\right)B/2+2A}\leq\left(C-2\sqrt{v}r\right)^{1/2}$
for some constant $C$ at large $r$. Again it means that $e^{\left(d-2\right)B/2+2A}$
must vanish at some finite $r$.

If we are under the second assumption on brane curvature, we can just
change $v_{0}e^{-2\left(d-2\right)B/2-8A}$ to $v_{0}e^{-2A}$ in
Eq. (\ref{eq:Jbound}). Since we have $e^{-2A}\geq e^{-\left(d-2\right)B/2-2A}$
at large $r$, the same proof with different numerical coefficients
applies.

Finally, we consider a bulk which is product of $d_{c}$ compact extra
dimensions and $d_{nc}\geq2$ noncompact extra dimensions. The brane
has codimension $\left(d_{c}+d_{nc}\right)$. Schematically, let the
metric be

\begin{equation}
ds^{2}=e^{2A\left(r\right)}\bar{g}_{\mu\nu}dx^{\mu}dx^{\nu}+dr^{2}+e^{2B\left(r\right)}\hat{g}_{ab}\left(\theta\right)d\theta^{a}d\theta^{b}+e^{2D\left(r\right)}ds_{C}^{2}\,,
\end{equation}

\noindent where $ds_{C}^{2}$ is the metric of the compact extra dimensions,
which does not depend on $x,r,\theta$. Now $a,b$ run from $1$ to
$\left(d_{nc}-1\right)$.To specify the compactness, we let $e^{2D}$
be finite and small for all $r$, while $e^{2B}$ diverges to infinity.
It is natural that $D$ only contributes to the field equations subdominantly
at large $r$ when comparing to $B$. Therefore the large $r$ behavior
of the system is not affected and the above proof still holds. This
finishes the proof of the desired no-go theorem.

\section{Conclusion\label{sec:Conclusion}}

We have studied the invalidity of the brane world scenario with noncompact
extra dimensions when compared to our observation of four-dimensional
gravity. That is, we show that localization of gravity around the
brane is not achievable within our general model, which is codimension-2
branes in at least one noncompact extra dimension, or higher codimensional
branes in at least two noncompact extra dimensions. We therefore conclude
that compactification is necessary to build consistent extra dimensions
in such setup.

Below we add some remarks to our results and point out some future
directions:
\begin{itemize}
\item In literature, we already know that noncompact extra dimensions are
usually not favorable for phenomenology. The point of our result is
that we give quantitative statements to formally exclude the possibility
of localized gravity in certain setups. We give explicit sufficient
conditions for the no-go theorems to hold. Therefore, the no-go theorems
are still more robust than the implicit statement in literature. 
\item We have proved the no-go theorem in a quite general context, but we
are not claiming that localization of gravity with two or more noncompact
extra dimensions must be impossible. It is easy to go beyond the no-go
theorem by, for example, adding higher-derivative terms into the action
and field equations, or consider different kinds of background such
as that with only the scalar field. Indeed, in string theory a $R^{4}$
coupling in noncompact bulk can induce 4D Einstein gravity on the
brane \cite{Antoniadis:2002tr}. This is a clear counterexample of
the no-go theorem. On the other hand, we are not saying that localization
of gravity becomes easy outside our conditions in the no-go theorem.
It remains interesting to understand how general the no-go theorem
can be.
\item We also exclude the case of codimension-2 branes with one compact
and one noncompact extra dimensions. The proof involves many properties
appear only in codimension-2 models, and it is non-trivial whether
the cases with more compact extra dimensions are also excluded. A
more general proof is therefore needed to complement our no-go theorem.
\item In the proof of our results, the techniques we used are almost purely
mathematical, except applying some basic physical properties of the
system. To understand more the models, a physical interpretation,
or even derivation of our results is needed. Especially, we should
have a clear understanding on how the distinct dynamics at brane position
between codimension-1 and higher-codimension models, which is introduced
in Section \ref{sec:Introduction}, physically abandon the possibility
of localized gravity. Surely symmetries play a crucial role in our
derivations, but it may be possible to construct the no-go theorem
by arguments based on only symmetries and topologies of the system,
but not referring to the explicit dynamics of the system. 
\end{itemize}
We hope to address some of the above issues in our further studies.

\noindent \vspace{4mm}

\section*{Acknowledgement}

The author thanks Cliff Burgess for initiating this work and initial
collaboration. He thanks Raman Sundrum, Andrew Tolley and Henry Tye
for useful discussions. He also thanks Ignatios Antoniadis for pointing
out their work as a counterexample. The bulk of this work was done
when the author visited Perimeter Institute. Research at Perimeter
Institute is supported in part by the Government of Canada through
the Department of Innovation, Science and Economic Development Canada
and by the Province of Ontario through the Ministry of Economic Development,
Job Creation and Trade.

\bibliography{refs}
 \bibliographystyle{JHEP}
\end{document}